\documentclass[letterpaper, 10 pt, conference]{ieeeconf} 

\usepackage{amsfonts,amssymb,amsmath}
\usepackage{graphicx,graphics,epsfig,color}
\usepackage{rgsMacros}
\usepackage{subcaption}
\usepackage{comment}
\usepackage{hyperref}

\let\labelindent\relax
\usepackage{enumitem}
\usepackage{balance}
\definecolor{olivegreen}{rgb}{0.14,0.29,0}

\IEEEoverridecommandlockouts                              
\newtheorem{exe}{Example}
\newtheorem{corol}{Corollary}
\newtheorem{ass}{Assumption}
\newtheorem{proper}{Property}
\newtheorem{defin}{Definition}
\newtheorem{prob}{Problem}
\newtheorem{cla}{Claim}
\newtheorem{rem}{Remark}
\newtheorem{lem}{Lemma}
\newtheorem{prop}{Proposition}
\newtheorem{thm}{Theorem}
\newtheorem{fct}{Fact}

\newenvironment{remark}{\begin{rem} \rm}{ \end{rem}}

\newenvironment{theorem}{\begin{thm}}{\hfill $\square$ \end{thm}}

\newif\ifitsdraft

\newtheorem{dwellt}{Condition}

\newif\ifitsdraft

\usepackage{mathtools}
\definecolor{cadmiumgreen}{rgb}{0.0, 0.42, 0.24}
\usepackage{tikz}
\usetikzlibrary{shapes,positioning,intersections,quotes}
\usepackage{pgfplots}
\pgfplotsset{compat=1.16}

\tikzset{
    cross/.pic = {
    \draw[rotate = 45] (-#1,0) -- (#1,0);
    \draw[rotate = 45] (0,-#1) -- (0, #1);
    }
}
\usepackage{amsfonts,amssymb,amsmath}

\usepackage{xcolor}

\usepackage{subcaption}

\newlength{\overwritelength}
\newlength{\minimumoverwritelength}
\newcommand{\overwrite}[3][red]{%
  \settowidth{\overwritelength}{$#2$}%
  \ifdim\overwritelength<\minimumoverwritelength%
    \setlength{\overwritelength}{\minimumoverwritelength}\fi%
  \stackrel
    {%
      \begin{minipage}{\overwritelength}%
        \color{#1}\centering\small #3\\%
        \rule{1pt}{9pt}%
      \end{minipage}}
    {\colorbox{#1!50}{\color{black}$\displaystyle#2$}}}

\usepackage{color}

\usepackage{pifont}
\usetikzlibrary{arrows.meta,positioning,calc,decorations.pathmorphing}

\tikzset{
  snakeline/.style = {->,thick, decorate, decoration={pre length=0.2cm, post length=0.2cm, snake, amplitude=.4mm, segment length=2mm}, cadmiumgreen},
  block/.style = {draw, fill=blue!20, minimum height=3em, minimum width=3em},
  pinstyle/.style={pin edge={to-,thin,black}},
}

\usetikzlibrary{shapes.misc}

\tikzset{cross/.style={cross out, draw=black, minimum size=2*(#1-\pgflinewidth), inner sep=0pt, outer sep=0pt},
cross/.default={1pt}}

\makeatletter
\newcommand \EmptyArgParse [2]
  {%
    \if\relax\detokenize{#2}\relax
      \def\ProcessedArgument{#1}%
    \else
      \def\ProcessedArgument{#2}%
    \fi
  }
\NewDocumentCommand \mathcircled { >{\EmptyArgParse{red}}O{} O{circle} m }
  {%
    \mathpalette{\mathcircled@b{#1}{#2}}{#3}%
  }
\newcommand\mathcircled@b[4]
  {%
    \tikz[baseline=(math.base)]
      \node[draw,color=#1,#2,inner sep=1pt] (math) {$\m@th#3#4$};%
  }
\makeatother

\title{\LARGE \bf 
From Sontag's to Cardano-Lyapunov Formula for  Systems\\ Not Affine in the Control: Convection-Enabled PDE Stabilization
}

\author{M. C. Belhadjoudja, M. Krsti{\'c}, M. Maghenem, and E. Witrant 
\thanks{M. C. Belhadjoudja and M. Maghenem are with Universit\'e Grenoble Alpes, CNRS, Grenoble-INP, GIPSA-lab, F-38000, Grenoble, France (e-mail: mohamed.belhadjoudja@gipsa-lab.fr).}
\thanks{E. Witrant is with Universit\'e Grenoble Alpes, CNRS, Grenoble-INP, GIPSA-lab, F-38000, Grenoble, France, and the Departement of Mechanical Engineering, Dalhousie University, Halifax B3H 4R2, Nova Scotia, Canada.}
\thanks{
M. Krsti{\'c} is with the Department of Mechanical and Aerospace Engineering, University of California San Diego, 92093
San Diego, USA (e-mail: krstic@ucsd.edu).}
\thanks{The project has been partially supported by CNRS, \textit{appel unique}, 2022.}}

\begin{document}

\maketitle

\begin{abstract}
    We propose the first generalization of Sontag's universal controller to systems not affine in the control, particularly, to PDEs with boundary actuation. We assume that the system admits a control Lyapunov function (CLF) whose derivative, rather than being affine in the control, has either a depressed cubic, quadratic, or depressed quartic dependence on the control. For each case, a continuous universal controller that vanishes at the origin and achieves global exponential stability is derived. We prove our result in the context of convection-reaction-diffusion PDEs with Dirichlet actuation. We show that if the convection has a certain structure, then the $L^2$ norm of the state is a CLF.
    %For flow convection $\pm (u^2)_{x}$, the CLF has a depressed cubic structure in the control input. For counter-convection $u_{x}$, the CLF has a quadratic structure. Finally, for Buckmaster convection $(u^3)_{x}$, the CLF has a depressed quartic structure. For each convection, stabilization is achieved for any reaction.
    In addition to generalizing Sontag's formula to some non-affine systems, we present the first general Lyapunov approach for boundary control of nonlinear PDEs. 
    %The only general method of nonlinear ODE control that has been extended to nonlinear PDEs is backstepping but the resulting controller is incredibly complex, as it entails infinite Volterra series and estimates for PDEs defined on Goursat domains with a dimension that increases to infinity but a volume that goes to zero.
    %Our controllers are extraordinarily simple in comparison to nonlinear backstepping, as they involve only a few standard algebraic operations.
    We illustrate our results via a numerical example.
    %We illustrate our results via a numerical example. where our universal controller for flow convection $-(u^2)_{x}$ is used to stabilize a PDE that involves the superlinear reaction term $u^3$, which leads to finite time blow-up in open-loop. 
\end{abstract}

\section{Introduction and Main Result}\label{introduction}

\subsection{Sontag's formula}
Artstein considers in his 1983 paper \cite{artstein} the problem of smooth stabilization of control systems
\begin{align}
%   \frac{\rm d}{{\rm d}t} 
   \dot u = f(u,v),
\end{align}
where, for consistency with the notation used in PDE control, $u$ denotes the state, $v$ denotes the control input, and $f$ a smooth vector field. One of Artstein's results is that, if the system admits a smooth CLF,  then there exists a stabilizing feedback  $v(u)$ with $v(0)=0$, which is smooth everywhere except possibly at the origin $u=0$. If the system possesses the `small control property,' then continuity at the origin can be guaranteed. 

Artstein's proof is not constructive: the feedback is not designed. For systems affine in the control, 
\begin{align}
%     \frac{\partial}{\partial t}
     \dot u = f(u)+g(u) v, \label{affine}
\end{align}
where $f$ and $g$ are smooth, Sontag provides a feedback formula in his 1989 paper \cite{sontag_universal}. Among various pedagogical presentations of Sontag's universal controller, the reader may consult Section 1.3 in \cite{krstic_uncertain}. 

Extensions of Sontag's formula are too numerous to comprehensively survey. For example, in \cite{sontag_bounded}, a bounded version of Sontag's controller is proposed under the assumption that the system admits a CLF whose derivative can be made negative by choosing the control input $v$ in a bounded set. In \cite[equation (2.4)]{krstic_sontag}, the concept of adaptive CLF is introduced for stabilization of nonlinear systems, linear in the unknown constant parameters, and an ``adaptive Sontag's formula'' is given. The extension of Sontag's formula to systems affine in a deterministic disturbance is given in \cite[equation (35)]{krstic1998inverse}. Sontag's formula for stochastic systems is introduced in \cite[equation (5.12)]{krstic_stochastic_sontag}. In \cite{iss_sontag}, Sontag's formula is generalized to the case where the input is affected by a disturbance, to guarantee integral-input-to-state stability. Finally, in \cite{sontag_event_based}, Sontag's formula is adapted for the problem of event-based stabilization of control-affine systems. 

\subsection{Going beyond control-affine systems: Framework}

None of the results above go beyond the control-affine case. %Sontag's formula has been adapted to various scenarios and has been improved by many authors, but the fundamental structure of the CLF in all those references remains the same. Namely, a control-affine CLF.
Our first contribution here is to take the construction of a universal feedback controller beyond the control-affine case, motivated by classes of boundary-actuated PDEs. Specifically, the question we ask, a third of a century after \cite{sontag_universal}, is: Given a CLF $V$ for a control system (not necessarily finite-dimensional), for which $\dot{V}$ is not affine in the control $v$, can we construct a 'universal' feedback controller $v$ that vanishes at the origin, is at least 'continuous everywhere', and stabilizes the existing solutions? In other words, can we extend the universal control concept beyond the control-affine case? 

To answer the above question, we consider the  class of scalar-valued PDE systems admitting a CLF $V$, whose time derivative has a polynomial structure in the control input $v$, i.e., 
\begin{align}
    \dot{V}(u) \leq \phi (u)+\gamma (v,u),\label{non-affine}
\end{align}
where $\phi $ is some operator acting on the state $u$ that verifies $\phi (0)=0$ and such that $\phi (u)$ does not depend explicitly on $v$, and $\gamma (v,u)$ is a polynomial in the control input $v$, whose higher-order coefficient is constant, and whose lower-order coefficients may depend on $u$ but not on $v$. 

The inspiration for this work comes from \cite{krsticBurgers1999}, \cite{liu2001stability}, \cite{KS1}, \cite{ACC23_KS}, and \cite{non-smooth1}. In \cite{krsticBurgers1999}, stabilization of the inviscid Burgers' equation is achieved by exploiting the quadratic convection $-(u^2)_{x}$, which makes $\dot{V}$ cubic in the control, with $V$ being the $L^2$ norm of the state. In \cite{liu2001stability}, \cite{KS1}, and \cite{ACC23_KS}, it is shown that the $L^2$ norm of the state of the Kuramoto-Sivashinsky is a CLF, because of the convection $-(u^2)_{x}$. Stabilization is then achieved. In \cite{non-smooth1}, the authors consider a reaction-diffusion PDE, observe that a spatially weighted $L^2$ norm of the state is quadratic in the control, thanks to the diffusion term, and propose a stabilizing feedback, which is a root of a quadratic equation.

% In this paper,  we rely on convection, in the parabolic case, for an (unweighted) $L^2$ norm CLF. 
We consider the following structures for $\gamma$: 
\begin{align}
&\gamma (v,u) := \pm v^3+\beta(u) v \quad &\text{(depressed cubic),} \label{control1}\\
&\gamma (v,u) := -v^2 + \beta(u) v \quad &\text{(quadratic),} \label{control2}\\
&\gamma (v,u) := -v^4+\beta (u)v \quad &\text{(depressed quartic),} \label{control3}
\end{align}
where $\beta$ is some continuous operator acting on $u$ that verifies $\beta (0)=0$. For each structure on $\gamma (v,u)$, we construct a universal continuous feedback  
\begin{align}
    v\left(V(u),\beta (u),\phi (u)\right)\label{universal}
\end{align}
that verifies $v(0)=0$ and, under which, the regular closed-loop solutions verify 
\begin{equation}
    \dot{V}(u) \leq -\alpha( V(u))\,, \quad \alpha\in {\cal K}\, , \label{I3}
\end{equation}
where $\alpha \in \mathcal{K}$ if $\alpha $ is continuous, strictly increasing and $\alpha (0)=0$. For example, one can choose $\alpha(V):=V$ to enforce exponential stability. The controller \eqref{universal} is said to be universal because its structure is not specific to $\phi (u)$ and $\beta (u)$, which are general/universal. The structure of the controller depends only on the structure of the {actuated part of the CLF}, i.e. $\gamma (v,u)$.

The universal feedback \eqref{universal}, when $\gamma (v,u)$ has the cubic structure \eqref{control1}; namely, when 
\begin{equation}\label{eq-Vdot-cubic}
    \dot V \leq \phi +\beta v \pm v^3 \,,
\end{equation}
is given by what we suggest to be called the {\em Cardano-Lyapunov formula} 
\begin{align}
\fbox{$\displaystyle v := \sqrt[3]{-\frac{q}{2}+\sqrt{\frac{q^2}{4}\pm\frac{\beta^3}{27}}}+\sqrt[3]{-\frac{q}{2}-\sqrt{\frac{q^2}{4}\pm \frac{\beta^3}{27}}}$} \label{Cardano_Lyapunov}
\end{align}
where %$p$ and $q$ are
\begin{align}
%\beta :=&~ \beta (u), \\
\fbox{$\displaystyle q := ~ \pm \left[|\phi (u)|+ \frac{2\sqrt{3}}{9}|\beta (u)|^{\frac{3}{2}}+\alpha (V(u))\right]$}
\end{align}
Note that, unlike the classical control-affine case, $\dot V = L_fV + L_gV v$, where one has to assume that $L_gV = 0 \Rightarrow L_f V<0$, in \eqref{eq-Vdot-cubic} the cubic term $\pm v^3$ removes the need for any assumptions on $\phi$ and $\beta$---namely, $V$ is a CLF by virtue of  \eqref{eq-Vdot-cubic} alone.

%The subtle difference between the two is the building block of universal control theory. To understand our idea, let us go back to Sontag's formula. A CLF for system \eqref{affine} satisfies 
%\begin{align}
 %   \dot{V}(u) = L_{f}V(u)+L_{g}V(u) v. 
%\end{align}
%For this particular structure on $\dot{V}$, we have a universal controller which is Sontag's formula. The structure of the universal controller is the same for any $L_{f}V(u)$ and $L_{g}V(u)$. These two functions are just arguments of the universal controller. By changing the structure of the stabilizing part of the CLF, i.e. by changing the way that the control input acts on the system, we change the universal controller. For example, if our system admits a CLF of the form 
%\begin{align}
 %   \dot{V}\leq f(u)-v^3+\beta (u)v,
%\end{align}
%then there exists a universal controller, which is not Sontag's formula, and whose structure is independent of the functions $f(u)$ and $\beta (u)$. The structure of the controller depends only on how $v$ acts on $\dot{V}$. To construct the universal controller we do not need to know what is $\beta (u)$. We just need to know that $\dot{V}$ is a depressed cubic in the control input. The coefficients $f(u)$ and $\beta(u)$ are just arguments of the controller. 

\subsection{Going beyond control-affine systems: Motivation}

We illustrate our result on the class of scalar convection-reaction-diffusion (CRD) PDEs of the form
\begin{align}
u_{t} = \epsilon u_{xx} + C(u)_{x}+R(u)\, \qquad x \in (0,1),\label{eq1}
\end{align}
where $R(u)$ is the reaction which is continuous and verifies $R(0)=0$; $C(u)_{x}$ is the convection; and $\epsilon u_{xx}$ is the diffusion with $\epsilon >0$. We suppose that this system is subject to the Dirichlet actuation 
\begin{eqnarray}
    &u(0) = v, \label{eq2-1}\\
    &u(1) = 0, \label{eq2-2}
\end{eqnarray}
in which $v$ is the control input to be designed to stabilize the origin $u:=0$ of \eqref{eq1} in the $L^2$ sense. Three types of convection are considered; namely, 
\begin{align}
&C(u)_{x} := \pm (u^2)_{x} \quad &\text{(flow convection),}\label{flow} \\
&C(u)_{x} := u_{x} \quad &\text{(counter-convection),} \label{linear}\\
&C(u)_{x} := (u^3)_{x} \quad &\text{(Buckmaster convection).}\label{buckmaster}
\end{align}
We reveal that the $L^2$ norm 
\begin{align}
    V(u) := \frac{1}{2}\int_{0}^{1}u(x)^2dx\label{eqdefV}
\end{align}
is a CLF for the control system \eqref{eq1}, under \eqref{eq2-1}, \eqref{eq2-2}, if the convection is given by \eqref{flow}, \eqref{linear}, or \eqref{buckmaster}. In particular, $\dot{V}$ is not affine in the control input $v$ and satisfies \eqref{non-affine}, with  $\gamma (v,u)$ a depressed cubic in the case of flow convection; quadratic in the case of counter-convection; and depressed quartic in the case of Buckmaster convection. The function $\phi (u)$ in \eqref{non-affine} depends only on the reaction and the diffusion, while the function $\beta(u)$ in \eqref{control1}, \eqref{control2} and \eqref{control3} comes only from the diffusion. 

CRD PDEs are difficult to control because of the reaction, which is the main source of instability. An example of a CRD PDE that is difficult %(if not impossible) 
to stabilize is  
\begin{align}
u_{t} = \epsilon u_{xx} - (u^2)_{x} +\lambda u^3\label{blow_up}
\end{align}
where $\lambda >0$ is constant. The superlinear reaction $u^3$ contributes to blow-up type phenomena in open-loop and may lead to the lack of global controllability \cite{superlinear1,superlinear2,superlinear3}.  Add to that, standard approaches such as linear backstepping cannot be applied to this system. For some PDEs involving superlinear reaction terms, it is possible to use nonlinear backstepping \cite{Volterra3,volterra_design,volterra_analysis}, but the resulting controller is extremely complex. It involves infinite Volterra series and the analysis of PDEs satisfied by the Volterra kernels, that evolve on Goursat domains. The dimension of the domains increases to infinity but the volume decreases to zero. 

\subsection{Main result}

For each case of the convection term $C(u)_x$, we design a feedback controller that achieves stabilization in the sense of inequality \eqref{I3}. Note that our approach is not limited to the CRD PDEs. It can be easily adapted to any control system that admits a CLF whose derivative verifies \eqref{non-affine}, under one of the structures \eqref{control1}, \eqref{control2}, or \eqref{control3}.

\begin{theorem}
\label{theorem1}
Consider the control system \eqref{eq1}, \eqref{eq2-1}, \eqref{eq2-2}, along with $V$  defined in \eqref{eqdefV}. For each of the convection terms in \eqref{flow}, \eqref{linear}, and  \eqref{buckmaster}, there exists a continuous feedback  
\begin{align}
v\left(V,u_{x}(0),\int_{0}^{1}\left(u(x)R(u(x))-\epsilon u_{x}(x)^2\right)dx\right), 
\label{feedback}
\end{align}
with the property $v(0)=0$ and guaranteeing that \eqref{I3} holds for the closed-loop regular solutions. 
\end{theorem}
The rest of the paper is devoted to the proof of Theorem \ref{theorem1}, i.e. to the construction of the feedback \eqref{feedback}. The universal controllers that we construct in this paper are remarkably simple in comparison with nonlinear backstepping controllers, as they are constructed using only finite number of additions, multiplications, and extraction of $n^{\rm th}$ roots. Hence, our result not only generalizes Sontag formula to non-affine cases, but also introduces a new method for boundary control of PDEs, distinct from PDE backstepping. We just need the PDE to possess a \textit{helpful convective structure}.
 
\section{Construction of the Universal Controllers}
In this section, we provide a constructive proof of Theorem \ref{theorem1}, i.e. we design the feedback \eqref{feedback}. The proof follows in two steps. First, we differentiate the Lyapunov function candidate $V$ along the regular solutions of \eqref{eq1}, \eqref{eq2-1}, \eqref{eq2-2}. We obtain, on $\dot{V}$, a structure that is not affine in the control, in the form of \eqref{non-affine}. Then, we consider the cases of flow convection $\pm (u^2)_{x}$, counter-convection $u_{x}$, and Buckmaster convection $(u^3)_{x}$. We show that the flow convection leads to a depressed cubic structure on the control input $v$, the counter-convection leads to a quadratic structure, and that the Buckmaster convection leads to a depressed quartic structure. In each case, we construct the corresponding universal controller. 

\subsection{Upperbounding $\dot{V}$} 
By differentiating $V$ along the regular solutions of \eqref{eq1}, \eqref{eq2-1}, \eqref{eq2-2}, we find
\begin{align}
\dot{V} =&~ \int_{0}^{1}u(x)C(u(x))_{x}dx +\int_{0}^{1}u(x)R(u(x))dx \nonumber \\
&~+\epsilon \int_{0}^{1}u(x)u_{xx}(x)dx. \label{new1}
\end{align}
Using integration by parts and the boundary conditions \eqref{eq2-1}, \eqref{eq2-2}, we obtain
%, we can rewrite \eqref{new1} as follows
%\begin{align}
%\dot{V} =&~ \bigg[u(x)C(u(x))\bigg]_{x=0}^{x=1}-\int_{0}^{1}u_{x}(x)C(u(x))dx \nonumber \\
%&~+\int_{0}^{1}u(x)R(u(x))dx + \epsilon \bigg[u(x)u_{x}(x)\bigg]_{x=0}^{x=1} \nonumber \\
%&~-\epsilon \int_{0}^{1}u_{x}(x)^2dx. \label{dirichlet1}
%\end{align}
%Using the boundary conditions \eqref{eq2-1}, \eqref{eq2-2}, we obtain
\begin{align}
&\dot{V} = -vC(v) - \int_{0}^{1}u_{x}(x)C(u(x))dx -\epsilon vu_{x}(0)\nonumber \\
&~ -\epsilon \int_{0}^{1}u_{x}(x)^2dx+\int_{0}^{1}u(x)R(u(x))dx. \label{dirichlet2}
\end{align}
Next, we perform the change of variable $s := u$, leading to $ds = u_{x}dx$. Equation \eqref{dirichlet2} can therefore be rewritten as 
\begin{align}
\dot{V} =&~ -vC(v) + \int_{0}^{v}C(s)ds -\epsilon \int_{0}^{1}u_{x}(x)^2dx\nonumber \\
&~+\int_{0}^{1}u(x)R(u(x))dx -\epsilon vu_{x}(0). \label{dirichlet3}
\end{align}
\subsection{Flow Convection}
We distinguish between two cases. Either $C(u):= u^2$, or $C(u) := -u^2$. Let us start with a $C(u):=u^2$. Note that
\begin{align}
-vC(v) + \int_{0}^{v}C(s)ds =&~-\frac{2}{3}v^3. 
\end{align}
As a consequence, equation \eqref{dirichlet3} becomes 
\begin{align}
\dot{V} =&~ \int_{0}^{1}u(x)R(u(x))dx-\epsilon \int_{0}^{1}u_{x}(x)^2dx \nonumber \\
&~-\frac{2}{3}v^3-\epsilon v u_{x}(0). \label{I6}
\end{align}
The function $\dot{V}$ has a depressed cubic structure on the control input $v$. The first idea to enforce \eqref{I3} is, therefore, to set $v$ as a real root of the cubic equation 
\begin{align}
&0 = v^3+\frac{3\epsilon}{2} v u_{x}(0) \nonumber \\
&~-\frac{3}{2}\left(\alpha (V) + \int_{0}^{1}\left(u(x)R(u(x)) - \epsilon u_{x}(x)^2\right)dx\right). \label{I7}
\end{align}
Selecting arbitrarily one real root of \eqref{I7} at each time instant may lead to $v$ being discontinuous with respect to the coefficients of \eqref{I7}. We need more work to enforce the continuity of $v$. This being said, we propose to consider instead the cubic equation
\begin{align}
0 =&~ v^3+\frac{3\epsilon }{2}u_{x}(0)v -\frac{3}{2}\bigg(\alpha (V) +\frac{\sqrt{2\epsilon^3}}{3}|u_{x}(0)|^{\frac{3}{2}} \nonumber \\
&~+\bigg| \int_{0}^{1}\left(u(x)R(u(x))-\epsilon u_{x}(x)^2\right)dx \bigg|\bigg). \label{I10}
\end{align}
If $v$ is a real root of \eqref{I10}, then we obtain 
\begin{align}
\dot{V} \leq&~ -\alpha (V) - \frac{\sqrt{2\epsilon^3}}{3}|u_{x}(0)|^{\frac{3}{2}} \nonumber \\
&~+\int_{0}^{1}\left(u(x)R(u(x))-\epsilon u_{x}(x)^2\right)dx\nonumber \\
&~ - \bigg|\int_{0}^{1}\left(u(x)R(u(x))-\epsilon u_{x}(x)^2\right)dx\bigg|, \label{I11}
\end{align}
which implies \eqref{I3}. It remains to choose the real root $v$ of \eqref{I10} such that it is continuous with respect to the coefficients of \eqref{I10}. To do so, we just need to remark that contrarily to \eqref{I7}, the depressed cubic equation \eqref{I10} always admits a real root, which is given by Cardano formula, which is a continuous expression relating the coefficients of \eqref{I10}. This is due to the fact that the discriminant of \eqref{I10} is nonnegative. Indeed, the discriminant of \eqref{I10} is $\Delta := q^2/4+\beta^3/27$, where $\beta$ and $q$ are given by 
\begin{align}
    \beta :=&~ \frac{3\epsilon}{2} u_{x}(0), \label{p1} \\
    q :=&~ -\frac{3}{2}\bigg(\alpha (V) + \bigg|\int_{0}^{1}\left(uR(u)-\epsilon u_{x}^2\right)dx\bigg| \bigg)  \nonumber \\
    &~ -\frac{\sqrt{2\epsilon^3}}{2}|u_{x}(0)|^{\frac{3}{2}}. \label{q1}
\end{align}
Note that 
\begin{align}
q =&~ -\frac{3}{2}\bigg(\alpha (V) + \bigg| \int_{0}^{1}\left(uR(u)-\epsilon u_{x}^2\right)dx\bigg| \bigg)  \nonumber \\
    &~ -\frac{2\sqrt{3}}{9}|\beta |^{\frac{3}{2}}. \label{q11}
\end{align}
A direct computation gives us 
\begin{align}
\Delta
    =&~ \frac{1}{4}\bigg(\frac{9}{4}\bigg(\alpha (V) + \bigg| \int_{0}^{1}\left(uR(u)-\epsilon u_{x}^2\right)dx\bigg| \bigg)^2\nonumber \\
    &~+\frac{\sqrt{3}}{3}\bigg(\alpha (V) + \bigg|\int_{0}^{1}\left(uR(u)-\epsilon u_{x}^2\right)dx\bigg| \bigg)\nonumber \\
    &~|\beta |^{\frac{3}{2}}+\frac{4}{27}|\beta |^3\bigg)+\frac{\beta^3}{27} \geq \frac{1}{27}(|\beta |^3+\beta^3) \geq 0. \label{delta1compu}
\end{align}
We, therefore, propose to set the control input $v$ as 
\begin{align}
    v:=\sqrt[3]{-\frac{q}{2}+\sqrt{\frac{q^2}{4}+\frac{\beta^3}{27}}}+\sqrt[3]{-\frac{q}{2}-\sqrt{\frac{q^2}{4}+\frac{\beta^3}{27}}}. \label{Cardano}
\end{align}
This control input vanishes if the state vanishes. Indeed, if $u=0$, then $\beta =q=0$,  which implies that $v=0$. 

Considering now a negative convection, we obtain
\begin{align}
-vC(v) + \int_{0}^{v}C(s)ds =&~ \frac{2}{3}v^3. 
\end{align}
As a consequence, equation \eqref{dirichlet3} becomes 
\begin{align}
\dot{V} =&~ \int_{0}^{1}u(x)R(u(x))dx-\epsilon \int_{0}^{1}u_{x}(x)^2dx \nonumber \\
&~+\frac{2}{3}v^3-\epsilon v u_{x}(0). \label{Inew}
\end{align}
To enforce inequality \eqref{I3}, we consider the cubic equation 
\begin{align}
0 =&~ v^3-\frac{3}{2}\epsilon u_{x}(0)v + \frac{3}{2}\bigg(\alpha (V) +\frac{\sqrt{2\epsilon^3}}{3}|u_{x}(0)|^{\frac{3}{2}}\nonumber \\
&~ +\bigg| \int_{0}^{1}\left(u(x)R(u(x))-\epsilon u_{x}(x)^2\right)dx \bigg|\bigg). \label{I10new}
\end{align}
The discriminant of \eqref{I10new} is $\Delta := q^2/4+\beta^3/27$ with $\beta$ and $q$ defined as 
\begin{align}
    \beta :=&~ -\frac{3\epsilon}{2} u_{x}(0), \label{p2} \\
    q :=&~ \frac{3}{2}\bigg(\alpha (V) + \bigg|\int_{0}^{1}\left(uR(u)-\epsilon u_{x}^2\right)dx\bigg| \bigg)  \nonumber \\
    &~ +\frac{\sqrt{2\epsilon^3}}{2}|u_{x}(0)|^{\frac{3}{2}}. \label{q2}
\end{align}
Since $\Delta \geq 0$, then we propose to set the control input $v$ as in \eqref{Cardano}. As previously, if $u=0$ then $\beta =q=0$ which leads to $v=0$. 
\subsection{Counter-convection}
In this case, we have 
\begin{align}
-vC(v)+\int_{0}^{v}C(s)ds
=&~ -\frac{1}{2}v^2. 
\end{align}
As a result, equation \eqref{dirichlet3} becomes 
\begin{align}
\dot{V} =&~ \int_{0}^{1}u(x)R(u(x))dx-\epsilon \int_{0}^{1}u_{x}(x)^2dx \nonumber \\
&~-\frac{1}{2}v^2-\epsilon v u_{x}(0). \label{I6new}
\end{align}
The function $\dot{V}$ has a quadratic structure in $v$. The first idea to enforce \eqref{I3}, is to set $v$ as a real root of the quadratic equation 
\begin{align}
0 =&~\int_{0}^{1}\left(u(x)R(u(x))-\epsilon u_{x}(x)^2\right)dx \nonumber \\
&~-\frac{1}{2}v^2-\epsilon u_{x}(0)v + \alpha (V).\label{39}
\end{align}
However, there is no guarantee that \eqref{39} admits a real root. We need to construct another polynomial equation as in the previous case. This being said, we consider the quadratic equation
\begin{align}
0 =&~ \bigg| \int_{0}^{1}\left(u(x)R(u(x))-\epsilon u_{x}(x)^2\right)dx \bigg| \nonumber \\
&~-\frac{1}{2}v^2-\epsilon u_{x}(0)v + \alpha (V).\label{40}
\end{align}
If $v$ is a real root of \eqref{40}, then \eqref{I3} is verified. The polynomial equation \eqref{40} admits either one double real root or two distinct real roots. This is due to the fact that its discriminant is nonnegative. Indeed, the discriminant of \eqref{40} is $\Delta := b^2-4ac$, where $a$, $b$ and $c$ are given by
\begin{align}
&a := -\frac{1}{2}, \quad b := -\epsilon u_{x}(0), \label{b1}\\
&c := \alpha (V)+\bigg| \int_{0}^{1}\left(uR(u)-\epsilon u_{x}^2\right)dx \bigg|.\label{c1}
\end{align}
Hence, $\Delta \geq 0$. We set then the control input $v$ as 
\begin{align}
    v := b\pm \sqrt{b^2+2c}. 
\end{align}
If $u=0$ then $b=c=0$ which implies that $v=0$.
\subsection{Buckmaster Convection}
The convection $(u^3)_{x}$ is characteristic of the Buckmaster equation \cite{buckmaster}. We have 
\begin{align}
-vC(v)+\int_{0}^{v}C(s)ds
=&~-\frac{3}{4}v^4. 
\end{align}
As a consequence, equation \eqref{dirichlet3} becomes 
\begin{align}
\dot{V} =&~ \int_{0}^{1}u(x)R(u(x))dx-\epsilon \int_{0}^{1}u_{x}(x)^2dx \nonumber \\
&~-\frac{3}{4}v^4-\epsilon v u_{x}(0). \label{I6new+}
\end{align}
The function $\dot{V}$ has a depressed quartic structure on $v$. The idea is to transform this depressed quartic structure into a biquadratic structure. Namely, using Young inequality, we find 
\begin{align}
    -\epsilon v u_{x}(0) \leq&~ \epsilon |v|\sqrt{|u_{x}(0)|}\sqrt{|u_{x}(0|} \nonumber \\
    \leq&~\frac{1}{2}|u_{x}(0)|v^2+\frac{\epsilon ^2}{2}|u_{x}(0)|.  
\end{align}
Equation \eqref{I6new+} becomes
\begin{align}
\dot{V} \leq&~ \int_{0}^{1}u(x)R(u(x))dx-\epsilon \int_{0}^{1}u_{x}(x)^2dx \nonumber \\
&~-\frac{3}{4}v^4+\frac{1}{2}|u_{x}(0)|v^2+\frac{\epsilon ^2}{2}|u_{x}(0)|.\label{final_new}
\end{align}
The function $\dot{V}$ has a biquadratic structure in $v$. We perform the change of variable $\tilde{v}:=v^2$, which allows us to rewrite \eqref{final_new} as 
\begin{align}
\dot{V} \leq&~ \int_{0}^{1}u(x)R(u(x))dx-\epsilon \int_{0}^{1}u_{x}(x)^2dx \nonumber \\
&~-\frac{3}{4}\tilde{v}^2+\frac{1}{2}|u_{x}(0)|\tilde{v}+\frac{\epsilon ^2}{2}|u_{x}(0)|.
\end{align}
To enforce inequality \eqref{I3}, we consider the quadratic equation 
\begin{align}
0 =&~ \bigg| \int_{0}^{1}\left(u(x)R(u(x))-\epsilon u_{x}(x)^2\right)dx \bigg| \nonumber \\
&~-\frac{3}{4}\tilde{v}^2+\frac{1}{2}|u_{x}(0)|\tilde{v} + \alpha (V)+\frac{\epsilon^2}{2}|u_{x}(0)|.\label{400}
\end{align}
The discriminant of \eqref{400} is given by  $\Delta := b^2-4ac$, with $a$, $b$ and $c$ defined as follows
\begin{align}
a :=&~ -\frac{3}{4}, \quad b := \frac{1}{2}|u_{x}(0)|, \label{b2}\\
c :=&~ \bigg| \int_{0}^{1}\left(u(x)R(u(x))-\epsilon u_{x}(x)^2\right)dx \bigg| \nonumber \\
&~+\alpha (V)+\frac{\epsilon^2}{2}|u_{x}(0)|.\label{c2}
\end{align}
Since the discriminant is nonnegative, we set the control input $v$ as
\begin{align}
    v := \pm \sqrt{\frac{2}{3}\left(b+\sqrt{b^2+3c}\right)}.
\end{align}
If $u=0$, then $b=c=0$ and therefore $v=0$. 
\begin{remark}
For the flow convection, we considered both $(u^2)_{x}$ and $-(u^2)_{x}$. For each direction, we constructed a universal controller. However, for the counter-convection and Buchmaster convection, only $u_x$ and $(u^3)_x$ are considered. To understand why, we consider for example the system
\begin{align}
    u_{t} = \epsilon u_{xx}-u_{x}+R(u).\label{negative_convection}
\end{align}
By differentiating $V$ along the regular solutions to \eqref{negative_convection}, \eqref{eq2-1}, \eqref{eq2-2}, and using integration by parts, we find 
\begin{align}
\dot{V} =&~ \int_{0}^{1}u(x)R(u(x))dx-\epsilon \int_{0}^{1}u_{x}(x)^2dx \nonumber \\
&~+\frac{1}{2}v^2-\epsilon vu_{x}(0). 
\end{align}
Hence, $V$ may not be a CLF. The same problem happens if we consider the system 
\begin{align}
u_{t} = \epsilon u_{xx}-(u^3)_{x}+R(u). \label{negative_Bconvection}
\end{align}
Indeed, by differentiating $V$ along the regular solutions to \eqref{negative_Bconvection}, \eqref{eq2-1}, \eqref{eq2-2}, and using integration by parts, we obtain
\begin{align}
    \dot{V} =&~ \int_{0}^{1}u(x)R(u(x))dx-\epsilon \int_{0}^{1}u_{x}(x)^2dx \nonumber \\
&~+\frac{3}{4}v^4-\epsilon vu_{x}(0).
\end{align}
Interestingly, however, our approach can be adapted to $-u_{x}$ and $-(u^3)_{x}$ if instead of the left Dirichlet actuation 
\eqref{eq2-1}-\eqref{eq2-2}, we consider the right Dirichlet actuation 
\begin{align}
u(0) = 0, \label{right_dirichlet1} \\
u(1) = v. \label{right_dirichlet2}
\end{align}
Indeed, if we differentiate now $V$ along the solutions to \eqref{negative_convection}, under  \eqref{right_dirichlet1}-\eqref{right_dirichlet2}, and we use integration by parts, we find 
\begin{align}
\dot{V} =&~ \int_{0}^{1}u(x)R(u(x))dx-\epsilon \int_{0}^{1}u_{x}(x)^2dx \nonumber \\
&~-\frac{1}{2}v^2+\epsilon vu_{x}(0).
\end{align}
%Similarly, by differentiation $V$ along the regular solutions to \eqref{negative_Bconvection}, \eqref{right_dirichlet1}, \eqref{right_dirichlet2}, and using integration by parts, we obtain 
%\begin{align}
 %   \dot{V} =&~ \int_{0}^{1}u(x)R(u(x))dx-\epsilon \int_{0}^{1}u_{x}(x)^2dx \nonumber \\
%&~-\frac{3}{4}v^4+\epsilon vu_{x}(0).
%\end{align}
In other words, $V$ is a CLF when we consider the convections $-u_{x}$ and $-(u^3)_{x}$ provided that the controller acts at the end point $x=1$ instead of $x=0$.
\end{remark}
\begin{remark}
Well-posedness of infinite dimensional systems such as \eqref{eq1} is far from being trivial, especially when the boundary controller is non-smooth. Non-smooth (yet continuous) feedback control of PDEs is not new. It first appears in \cite{krsticBurgers1999},  where a non-smooth boundary controller is designed to stabilize an inviscid Burgers' equation. However, since this first paper, except for very few works such as \cite{non-smooth1}, non-smooth control of PDEs has not been studied. The principal reason for such poor exploration is that it is difficult to prove closed-loop well-posedness via the standard functional analysis techniques used for more regular boundary controllers. This being said, well-posedness of infinite-dimensional systems subject to non-smooth (yet continuous) boundary feedback controllers is a fundamental mathematical question that has not been solved yet, but that should not hold us back from exploring what kind of stability results one can obtain using less regular controllers. 
\end{remark}
\section{Numerical Example}
In this Section, we illustrate our main result in Theorem \ref{theorem1} via a numerical example. The PDE we consider is \eqref{blow_up} under \eqref{eq2-1}, \eqref{eq2-2}. In open-loop ($v=0$), this PDE  exhibits finite time blow-ups due to the superlinear reaction term $u^3$ \cite{superlinear1,superlinear2,superlinear3}.

According to Theorem \ref{theorem1}, provided that for a given initial condition a regular closed-loop solution to \eqref{blow_up}, \eqref{eq2-1}, \eqref{eq2-2} exists, then this solution verifies \eqref{I3}. This being said, we will consider an initial condition for which a finite-time blow-up occurs in open-loop and show that our universal controller prevents such a blow-up and stabilizes the solution towards the origin. Since this equation possesses the flow convection $-(u^2)_x$, we select the control input $v$ as in \eqref{Cardano} with $\beta$ given by \eqref{p2} and $q$ given by \eqref{q2}. 
%\begin{align}
%q :=&~ \frac{3}{2}\bigg(\alpha (V) + \bigg|\lambda \int_{0}^{1}\left(u(x)^4-\epsilon u_{x}(x)^2\right)dx\bigg| \bigg)  \nonumber \\
 %   &~ +\frac{\sqrt{2\epsilon^3}}{2}|u_{x}(0)|^{\frac{3}{2}}.\label{qfinal}
%\end{align}

\begin{figure}
\centering % Not needed
\begin{subfigure}[b]{1\columnwidth}
\centering
        \includegraphics[width=\textwidth]{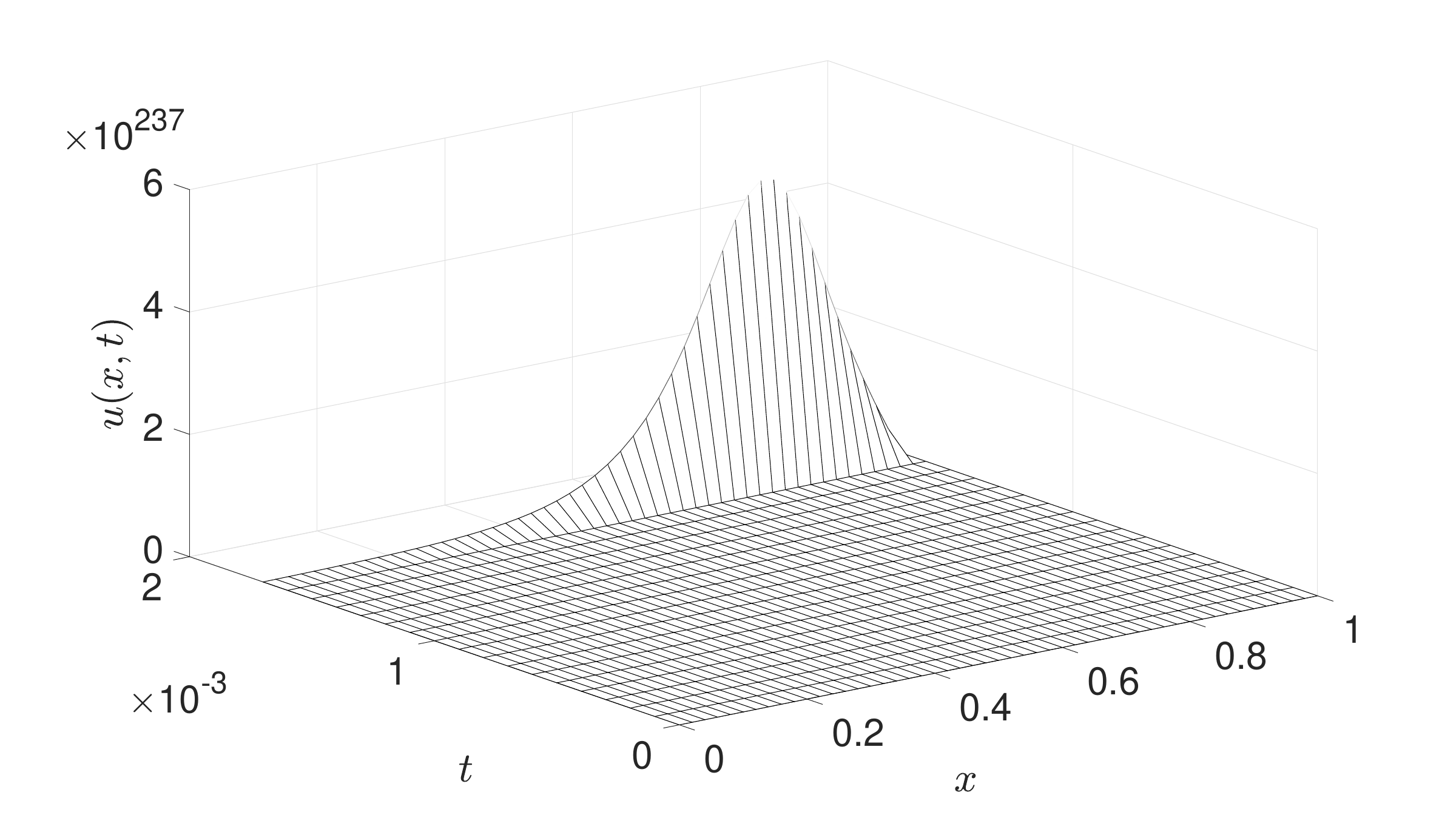}
        %\caption{Open-loop response.}
        \label{a}
    \end{subfigure}
    \hfill
    \begin{subfigure}[b]{1\columnwidth}
    \centering
        \includegraphics[width=\textwidth]{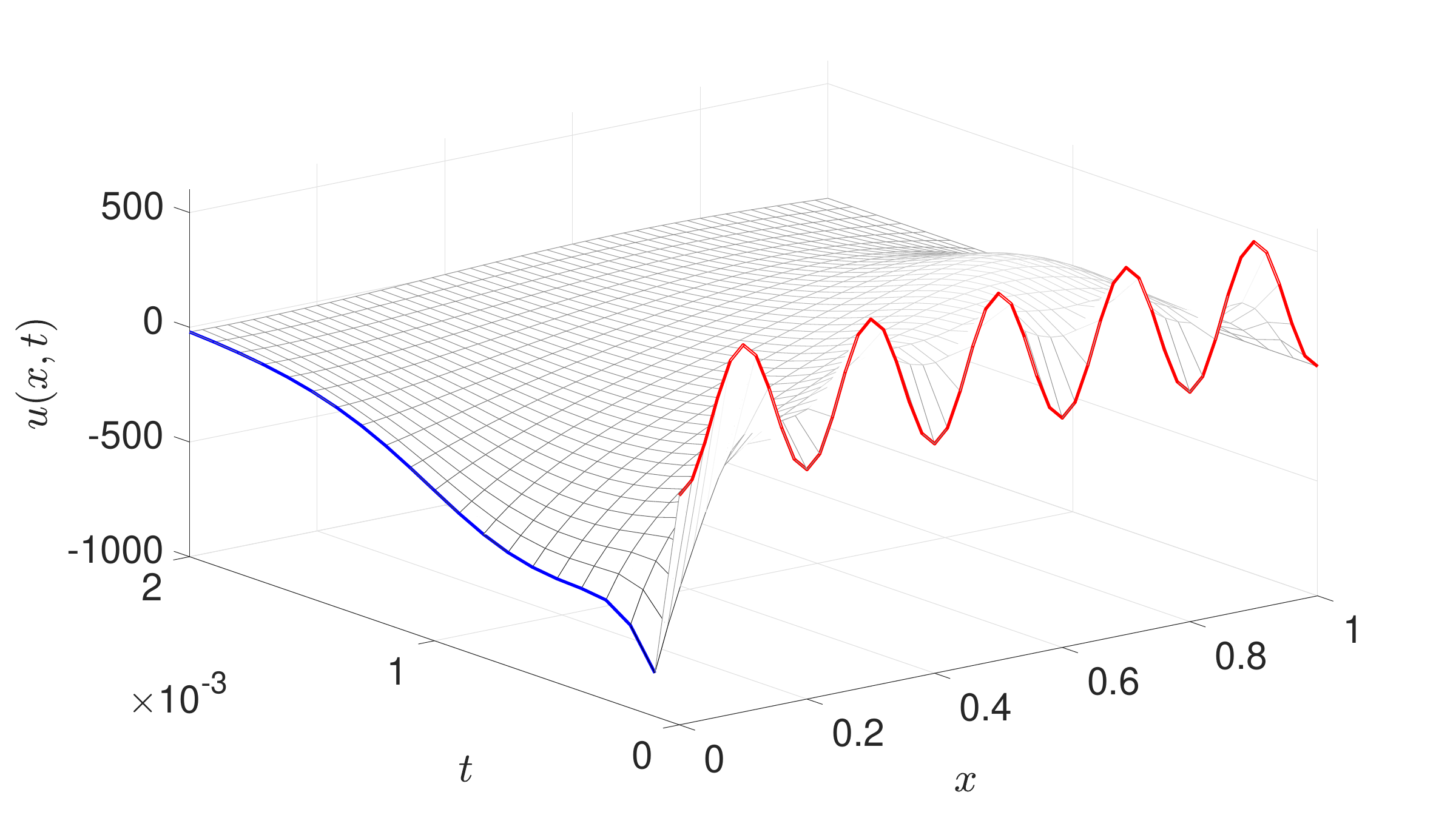}
        %\caption{Closed-loop response.}
        \label{b}
    \end{subfigure}
    %% leave a blank line to create a line break

    \caption{Open-loop response (top) of system \eqref{blow_up}. Closed-loop response (bottom) of system \eqref{blow_up}. %using the universal controller \eqref{Cardano}, \eqref{p2}, \eqref{qfinal}.
    }
    \label{fig1}
\end{figure}

For the simulations, we take $\alpha(V):=V$, and we select the coefficients $\lambda := 0.01$ and $\epsilon := 0.0002$. The initial condition is $u(x,0) := -300\left(\cos(10\pi x)-1\right)$. The simulations are performed on Matlab\textsuperscript{\textregistered} R2022b. We combine a Crank-Nicolson scheme with multiquadrics 
radial-basis-function (RBF) decomposition. We recall that multiquadrics RBF depend on a shape parameter $c\in \mathbb{R}$ that is to be chosen. We use an explicit scheme for the convection and the reaction. We use the same numerical parameters for both open-loop and closed-loop responses. Namely, the time step is $\Delta t := 10^{-4}$; the number of collocation points is $N+1:=501$; and the shape parameter is $c:=10^{-9}$. The open-loop response is depicted in Figure \ref{fig1} (top). We can observe the finite time blow-up of the solution. We close the loop with the universal controller \eqref{Cardano}, \eqref{p2}, \eqref{q2}. The closed-loop response is shown in Figure \ref{fig1} (bottom). We can see that our universal controller prevents finite-time blow-up, and stabilizes the solution. 
\section{Conclusion}
We proposed in this work the first generalization of Sontag's formula to systems that admit a CLF whose derivative is not affine in the control input, but polynomial. Depending on the structure of the resulting polynomial, we designed continuous universal controllers that vanish at the origin and achieve stabilization. We proved our results in the context of boundary control of convection-reaction-diffusion PDEs. Finally,
%The main motivation behind considering such structures on the derivative of a CLF comes from PDE boundary control. We consider CRD PDEs subject to Dirichlet actuation and show that the derivative of the $L^2$ norm of the state is a CLF of either cubic or quadratic structure depending on the convection. Our controllers represent a new method, distinct from PDE backstepping, to perform boundary control of certain PDEs, Namely, PDEs that possess \textit{helpful convection}.
our results have been illustrated on a numerical example where the PDE to control possesses a superlinear reaction term that yields finite-time blow-up phenomena. In future work, we would like to guarantee in addition to continuity boundedness of the boundary control input in the context of CRD PDEs. We would also like to see if it is possible to design universal controllers for other structures of the derivative of the CLF, and for more general convective terms, such as a combination of flow convection and counter-convection. It would be also interesting to generalize our approach to higher-order PDEs, such as the Kuramoto-Sivashinsky and the Korteweg-de Vries equations.
%namely a cubic reaction $u^3$ that leads to blow-up type phenomena in open-loop, and may even cause a lack of global controllability. The only technique available in the literature to deal with some PDEs involving superlinear reaction terms is nonlinear backstepping. However, the resulting controller is of extreme complexity. Our controllers are roots of polynomials solvable by radicals, as a consequence, they are obtained using only a finite number of the standard algebraic operations of addition, multiplication, and extraction of n$^{th}$ roots, which makes them remarkably simpler than the controllers obtained from nonlinear backstepping. 

\bibliography{biblio}
\bibliographystyle{ieeetr}

\end{document}